\documentclass [12 pt]{article}
\def\be{\begin{equation}}
\def\eea{\end{eqnarray}}
\def\bea{\begin{eqnarray}}
\def\ee{\end{equation}}

\usepackage[psamsfonts]{amssymb}
\usepackage{amsmath}

\author{M. Amooshahi$^{1}$
\footnote{Amooshahi@sci.ui.ac.ir}\footnote{Tel: 98-311-793-2418,\ \
Fax: 98-311-793-2409}
\\ $^{1}$ {\small Faculty of science, University of Isfahan ,Hezar Jarib Ave.,
Isfahan,Iran}}
\title{A Canonical Relativistic Approach to Quantize Electromagnetic Field in the Presence of Moving Magneto-Dielectric Media}
\begin{document}
\maketitle
\begin{abstract}
A canonical relativistic formulation is introduced to quantize
electromagnetic field in the presence of a polarizable and
magnetizable moving medium. The medium is modeled  by a continuum of
the second rank antisymmetric tensors in a phenomenological way. The
covariant wave equation for the vector potential and the covariant
constitutive equation of the medium are obtained as the Euler-
Lagrange equations using the Lagrangian of the total system. A
fourth rank tensor  which couples the electromagnetic field and the
medium is introduced. The susceptibility tensor of the medium is
obtained in terms of this coupling tensor. The noise polarization
tensor is calculated in terms of  both the coupling tensor and the
ladder
operators of the tensors modeling  the medium.
\end{abstract}
\section{Introduction}
 The electromagnetic field can be quantized in the presence of an absorbing  polarizable and
magnetizable medium by modeling such a medium by two independent
collections of oscillators\cite{1,2}. This method, in addition to
its simplicity and beauty, covers some general cases. This
quantization scheme has been generalized  for anisotropic media and
for spatially and temporarily dispersive media \cite{3,4}. It has
been shown that this formalism can be obtained canonically using a
Lagrangian function\cite{5}. The quantization method is also
applicable for a non-linear megnetodielectric medium.  One advantage
of this method is that, in contrast of the Huttner-Barnett model
\cite{6,7}, the electric and magnetic polarization fields are not
needed to be included in the Lagrangian of the total system as a
part of the degrees of freedom of the medium. That is, the
oscillators modeling the medium  constitute solely the degrees of
freedom of the medium and  are able to
describe both polarizability and absorptivity  of the medium.\\
\indent The electromagnetic field quantization in the presence of an
anisotropic dielectric media with spatio-temporal dispersion  has
also been treated previously  in the base of the quantum damped
polariton model \cite{8}.\\
\indent In the present work the previous model \cite{1}-\cite{5} has
been generalized for a moving medium  by using a relativistically
covariant approach. The medium is modeled , in a phenomenological
way,  by a continuum of the second rank antisymmetric tensors
labeled by a continuous parameter $\omega$. The tensors modeling the
medium describe both polarizability and absorptivity of the medium.
In order to represent   a canonical quantization in a
relativistically covariant language, we apply the lorentz gauge and
the Gupta-Bleuler method is used \cite{9}. In Gupta-Bleuler
quantization formalism the components of the four-vector potential
$A^\mu$ constitute the dynamical variables of electromagnetic field
and  only  those vectors , $|\psi\rangle$ in the
  Hilbert space of the total system, are admitted for which the expectation value of the gauge condition,  $ \langle\psi| \partial_\mu
A^\mu|\psi\rangle=0$, is satisfied. In this formalism  the Lorentz
condition can not be carried over to the field operators $A^\mu$ as
 $\partial_\mu A^\mu=0$, since the condition   $\partial_\mu
 A^\mu=0$  as an operator identity is incompatible to the
 commutation relations of the operator $A^\mu$ and its conjugate
 dynamical variable. \\
\indent For simplicity we set $c=1$ for the speed of light in the
vacuum and choose the space-time metric as $
g^{\mu\nu}=dia(-1,-1,-1,1)$.\\
\section{Relativistic classical electrodynamics in the presence a
magnetodielectric medium}
 In order to  have a classical treatment of  electrodynamics in the presence of a moving polarizable and magnetizable
 medium , the medium is modeled by a continium  of the second rank antisymmetric tensors
 $Y^{\alpha\beta}(\omega ,x)$ labeled by a continuous parameter $\omega$. The Lagrangian density
of the total system  is proposed as
\begin{eqnarray}\label{M1}
\pounds(x)&=& \frac{1}{2}\partial_\mu A_\nu\ \partial^\mu
A^\nu-\int_0^\infty d\omega \ f^{\mu\nu\alpha\beta}(\omega,x)\
\partial_\mu
A_\nu \ Y_{\alpha\beta}(\omega ,x)\nonumber\\
&+&\frac{1}{2}\int_0^\infty d\omega \left[
\partial_\mu Y_{\alpha\beta}(\omega ,x)\partial^\mu Y^{\alpha\beta} (\omega ,x)-\omega^2
Y_{\alpha\beta}(\omega ,x)Y^{\alpha\beta}(\omega ,x)\right]
\end{eqnarray}
where  $x=(t, \vec{x})$ stands for the coordinates of the space-time
and $ A^\mu$ is the four vector potential . Here the tensor
$f^{\mu\nu \alpha\beta}(\omega,x)$ is called the coupling tensor of
electromagnetic field and the medium and is antisymmetric with
respect to the first two indices $ \mu , \nu $ and also with respect
to the last two indices $\alpha , \beta$. Using the Lagrangian
(\ref{M1}), the Euler- Lagrange equations for the vector potential
are obtained as
\begin{equation}\label{M2}
\Box^2A^\nu=4\pi\
\partial_\mu K^{\mu\nu}
\end{equation}
where $\Box^2=\frac{\partial^2}{\partial t^2}-\nabla^2$ is the
d'Alembert's operator and
\begin{equation}\label{M3}
K^{\mu\nu}(x)=\frac{1}{4\pi}\int_0^\infty d\omega\ f^{\mu\nu
\alpha\beta}(\omega , x)Y_{\alpha\beta} (\omega , x)
\end{equation}
is the antisymmetric electric-magnetic polarization tensor of the
medium. Accordingly,   the polarization of the medium can be defined
naturally in terms of the antisymmetric tensors $Y^{\alpha\beta}
(\omega , x)$ modeling the medium and the polarization fields of the
medium are not needed  to be included explicitly  in the Lagrangian
of the total system. The electric-magnetic polarization tensor
$K^{\mu\nu}$ is related to the field strength tensor $
F^{\mu\nu}=\partial^\mu A^\nu-\partial^\nu A^\mu$ by  the covariant
form of the constitutive equation of the moving medium. Before
obtaining this equation from the Euler- Lagrange equations, it is
necessary  to point out the following  remarks:\\
\indent 1. The electric polarization density of a moving  medium
depends on both electric  and magnetic field , even though the
electric polarization is dependent only on the electric field when
the medium is at rest. This is so, since both the magnetic and
electric forces affect  the electric polarization of the  medium
when it moves. Similarly the magnetic polarization of a moving
medium in general may be dependent on both
electric and magnetic field\cite{10,11}.\\
\indent  2. A moving medium may be anisotropic even though the
medium is isotropic with respect to its rest frame\cite{12,13}. This
is clear, because the isotropy  of the medium is not preserved under
a Lorentz transformation. Furthermore the electromagnetic features
of the moving medium in direction of its motion are different from
those in
any other directions. \\
\indent 3. A  medium which is only spatially inhomogeneous with
respect to its rest frame is  both spatially and temporarily
inhomogeneous when it moves. Therefore a moving medium will be in
general both spatially and temporarily
dispersive  even if it can be assumed to respond locally  when it is at rest.\\
\indent To obtain the relativistically covariant form of the
constitutive equation of the medium which satisfies the  above
mentioned properties, we write the Euler-Lagrange equation for the
tensor $ Y^{\alpha\beta}(\omega , x)$
\begin{equation}\label{M4}
\Box^2Y^{\alpha\beta}(\omega,x)+\omega^2Y^{\alpha\beta}(\omega,x)= -
f^{\mu\nu\alpha\beta}(\omega , x)\partial_\mu A_\nu(x)
\end{equation}
By solving  this equation for the tensor $Y^{\alpha\beta}$ in terms
the retarded Green function \cite{11,14} and  substituting the
obtained solution in the definition of the polarization tensor
(\ref{M3}),  the covariant form of the constitutive equation of the
 the  medium relating the
polarization tensor $K^{\mu\nu}$ to the the electromagnetic field
strength tensor $F^{\mu\nu}$ is found as
\begin{equation}\label{M5}
K^{\mu\nu}(x)=K^{\mu\nu}_N(x)+\int_{-\infty}^{+\infty} d^4x'\
\chi^{\mu\nu\alpha\beta}(x,x')\ F_{\alpha\beta}(x')
\end{equation}
where $\chi^{\mu\nu\alpha\beta}(x,x')$ is the susceptibility tensor
of the medium which is antisymmetric with respect to the first two
indices $\mu , \nu$ and also with respect to the last two indices
$\alpha , \beta$ and is given by
\begin{eqnarray}\label{M6}
&&\chi^{\mu\nu\alpha\beta}(x,x')=\frac{1}{8\pi}\int_0^\infty
d\omega\ G( \omega, x-x') f^{\mu\nu\delta\sigma}(\omega,x)\
f^{\alpha\beta}_{\ \ \ \delta\sigma}(\omega,x')\nonumber\\
&=& - \frac{\vartheta(t-t')}{64\pi^4}\int_D\ d^4q\ \sin[ \
q\cdot(x-x')\ ]\ \frac{f^{\mu\nu \delta\sigma}(\sqrt{q^2},x)\
f^{\alpha\beta}_{\ \ \ \delta\sigma}(\sqrt{q^2},x') }{2\sqrt{q^2}}
\end{eqnarray}
Here $ \vartheta$ is the step function , the integration domain $ D$
is the upper half cone $ q_0^2-\vec{q}^2\geq 0\ ,\ q_0\geq 0 $  and
$ G( \omega, x-x')$ is the retarded Green function for Eq.
(\ref{M4}) satisfying
\begin{equation}\label{M7}
 \left[\Box^2
+\omega^2\right]G( \omega, x-x')=-\delta^{(4)}(x-x')
\end{equation}
In the constitutive equation (\ref{M5}), $K^{\mu\nu}_N$ is the noise
polarization tensor which is related to the solution of the
homogeneous equation $[\Box^2+\omega^2]Y^{\alpha\beta}_N=0$ as
follows
\begin{equation}\label{M8}
K^{\mu\nu}_N(x)=\frac{1}{4\pi}\int_0^\infty d\omega\ f^{\mu\nu
\delta\sigma}(\omega , x)Y_{N\delta\sigma} (\omega , x)
\end{equation}
From the solution of the wave equation (\ref{M4}) which is written
as
\begin{equation}\label{M8.1}
Y^{\mu\nu} (\omega , x)=Y_{N}^{\mu\nu} (\omega ,
x)+\int_{-\infty}^{+\infty} d^4x' G(\omega , x-x')\
f^{\mu\nu\alpha\beta}(\omega,x')\ \partial_\alpha A_\beta(x')
\end{equation}
since the retarded Green function $G( \omega, x-x')$ vanishes in the
limit $ t\rightarrow -\infty $
 it is clear that the noise fields
$Y_{N}^{\alpha\beta} (\omega , x)$ and  $K^{\mu\nu}_N(x)$ are
 the asymptotic forms of $Y^{\alpha\beta} (\omega ,
x)$ and $ K^{\mu\nu}$ for very large negative times, respectively.
The noise tensor field $Y_N^{\alpha\beta}(\omega,x)$ can be expanded
in terms of  the plane waves as follows
\begin{equation}\label{M8.1}
Y_N^{\alpha\beta}(\omega,x)=\frac{1}{\sqrt{(2\pi)^3}}
\int_{-\infty}^{+\infty} d^3k\ \left[
B^{\alpha\beta}(\omega,\vec{k})e^{\imath\vec{k}\cdot\vec{x}-\imath\sqrt{k^2+\omega^2}t}+
 C . C \right]
\end{equation}
In classical electrodynamics the tensor fields
$B^{\alpha\beta}(\omega,\vec{k})$
 are numerical functions which can be obtained in terms of
 the asymptotic forms of $Y^{\alpha\beta}(\omega,x)$ and $ \frac{\partial Y^{\alpha\beta}(\omega,x)}{\partial
 t}$  as
\begin{eqnarray}\label{M8.2}
&&B^{\alpha\beta}(\omega,\vec{k})\  e^{-\imath\sqrt{k^2+\omega^2}t}=\nonumber\\
&&\frac{1}{2 \sqrt{(2\pi)^3}}\int_{-\infty}^{+\infty} d^3x\
[Y_N^{\alpha\beta}(\omega,x)+  \frac{\imath}{\sqrt{k^2+\omega^2}}\
\frac{\partial Y_N^{\alpha\beta}(\omega,x)}{\partial
 t}]e^{-\imath\vec{k}\cdot\vec{x}}
\end{eqnarray}
In Quantum electrodynamics which is discussed in the next section
the  tensor fields $B^{\alpha\beta}(\omega,\vec{k})$ are operator
valued functions that act on the Fock space of the total system.
\section{Canonical quantization}
Having the Lagrangian of the total system given by (\ref{M1}), one
can compute the canonical conjugate momentum densities of the system
as
\begin{eqnarray}\label{M9}
-\Pi^\nu&=&\frac{\partial L}{\partial\left(\partial_0
A_\nu\right)}=\partial^0A^\nu-4\pi K^{0\nu}\nonumber\\
Q^{\alpha\beta}(\omega , x)&=&\frac{\delta
L}{\delta\left(\partial_0Y_{\alpha\beta}(\omega,x)\right)}=\partial^0Y^{\alpha\beta}(\omega,x)
\end{eqnarray}
According to the standard canonical quantization method, to quantize
the total system  the following equal-time commutation relations are
imposed on the dynamical fields $ A^\nu , Y^{\alpha\beta}(\omega,x)$
and their conjugate variables
\begin{equation}\label{M10}
\left[A^\mu(\vec{x},t)\ ,\ -\Pi^\nu(\vec{x'},t)\right]=\imath\hbar\
g^{\mu\nu}\ \delta^{(3)}(\vec{x}-\vec{x'})
\end{equation}
\begin{eqnarray}\label{M10.1}
 &&\left[Y^{\alpha\beta}(\omega,\vec{x},t)\
,\ Q^{\mu\nu}(\omega',\vec{x'},t)\right]=\imath\hbar\ [g^{\alpha\mu}
g^{\beta\nu}-g^{\alpha\nu} g^{\beta\mu}]
\delta(\omega-\omega')\delta^{(3)}(\vec{x}-\vec{x'})\nonumber\\
&&
\end{eqnarray}
The canonical momenta  (\ref{M9}) and the total Lagrangian
(\ref{M1}) can be used to write the Hamiltonian of the total system
as
\begin{eqnarray}\label{M11}
H(t)&=&\int_{-\infty}^{+\infty} d^3x\ \left[
\frac{\left(\Pi^\alpha-K^{0\alpha}\right)\left(\Pi_\alpha-K_{0\alpha}\right)}{2}\right]\nonumber\\
&+&\int_{-\infty}^{+\infty} d^3x\left[ -\frac{1}{2}\partial_l
A_\alpha\
\partial^l A^\alpha+4\pi\ \partial_l A_\alpha \ K^{l\alpha}\right]\nonumber\\
&+&\int_{-\infty}^{+\infty} d^3x\int_0^\infty d\omega\left[
\frac{1}{2}Q_{\alpha\beta}(\omega,\vec{x},t)
Q^{\alpha\beta}(\omega,\vec{x},t)\right]\nonumber\\
&-&\frac{1}{2}\int_{-\infty}^{+\infty} d^3x\left[\partial_l
Y_{\alpha\beta}(\omega,\vec{x},t)\partial^lY^{\alpha\beta}(\omega,\vec{x},t)\right]\nonumber\\
&+&\int_{-\infty}^{+\infty} d^3x\left[ \frac{1}{2}\omega^2 \
Y_{\alpha\beta}(\omega,\vec{x},t)Y^{\alpha\beta}(\omega,\vec{x},t)\right]
\end{eqnarray}
where the sum is done  over $\alpha$ from 0 to 3 and over $l$ from 1
to 3. It is easy to show that, in the Heisenberg picture, the
combination of the Heisenberg equations of the conjugate variables $
A^\nu$ and $-\Pi^\nu$ leads to the covariant wave equation
(\ref{M2}). Also using the Heisenberg equations of the conjugate
variables $ Y^{\alpha\beta}(\omega,x)$ and $
Q^{\alpha\beta}(\omega,x)$ gives us  Eq.(\ref{M4}). Now substituting
$K^{\mu\nu}$ from (\ref{M5}), into (\ref{M2}) reads to the equation
of motion of the vector potential as
\begin{equation}\label{M12}
\Box^2 A^\nu-8\pi \int_{-\infty}^{+\infty} d^4x'\ \partial_\mu
\chi^{\mu\nu\alpha\beta}(x,x')\ \partial_\alpha A_\beta (x')=4\pi\
\partial_\mu K^{\mu\nu}_N(x)
\end{equation}
By introducing  the following basis for the second rank
antisymmetric tensors
\begin{equation}\label{M13}
\eta^{\alpha\beta}(\vec{k},\lambda ,
\lambda')=\frac{\varepsilon^\alpha(\vec{k},\lambda)\varepsilon^\beta(\vec{k},\lambda')-\varepsilon^\alpha(\vec{k},\lambda')\varepsilon^\beta(\vec{k},\lambda)
}{\sqrt{2}}\hspace {1.00 cm}\lambda<\lambda',
\end{equation}
the noise tensor field $Y_N^{\alpha\beta}(\omega,x)$ can be expanded
in terms of  the plane waves as follows
\begin{eqnarray}\label{M14}
&&Y_N^{\alpha\beta}(\omega,x)=\sum_{\lambda,\lambda'=0}^3\int
_{-\infty}^{+\infty} d^3k\
\sqrt{\frac{\hbar}{4(2\pi)^3\sqrt{k^2+\omega^2}}}\times\nonumber\\
&&\left[
b_{\lambda\lambda'}(\omega,\vec{k})e^{\imath\vec{k}\cdot\vec{x}-\imath\sqrt{k^2+\omega^2}t}+h.c\right]\eta^{\alpha\beta}(\lambda,\lambda',\vec{k}),
\end{eqnarray}
where $\varepsilon^\alpha(\vec{k},\lambda)\hspace{.50
cm}\lambda=0,1,2,3$ \ are a basis for the four dimensional Minkowski
space and for any vector $\vec{k}$ obeying the orthonormality and
the completeness relations
\begin{eqnarray}\label{M15}
\varepsilon^\sigma(\lambda,\vec{k})\varepsilon_\sigma(\lambda',\vec{k})&=&g_{\lambda\lambda'}\nonumber\\
 \sum_{\lambda=0}^3
g_{\lambda\lambda'}\
\varepsilon_\sigma(\lambda,\vec{k})\varepsilon_{\sigma'}(\lambda,\vec{k})&=&g_{\sigma\sigma'}.
\end{eqnarray}
The  annihilation and creation operators
 $b_{\lambda\lambda'}(\omega,\vec{k})$ and
 $b^\dag_{\lambda\lambda'}(\omega,\vec{k})$ in the expansion (\ref{M14}) are antisymmetric with respect to
  the indices $ \lambda , \lambda'$. Since the noise fields $Y^{\alpha\beta}_N(\omega,x)$   and  $ \partial_0Y^{\alpha\beta}_N(\omega,x)$
  obey the same commutation relations (\ref{M10.1}),  the commutation relations between $b_{\lambda\lambda'}(\omega,\vec{k})$
  and $b^\dag_{\lambda\lambda'}(\omega,\vec{k})$ are
\begin{equation}\label{M16}
\left[b_{\lambda_1\lambda_2}(\omega,\vec{k})\ ,\
b^\dag_{\lambda'_1\lambda'_2}(\omega',\vec{k'})\right]=[g_{\lambda_1\lambda'_1}g_{\lambda_2\lambda'_2}-g_{\lambda_1\lambda'_2}g_{\lambda_2\lambda'_1}]\delta(\vec{k}-\vec{k'})\delta(\omega-\omega'),
\end{equation}
 where  the completeness and orthonormality relations
\begin{eqnarray}\label{M17}
&&\sum_{\lambda , \lambda'=0}^3\ g_{\lambda\lambda}\
g_{\lambda'\lambda'}\ \eta^{\alpha\beta}(\lambda , \lambda',\vec{k})
\eta^{\mu\nu}(\lambda , \lambda', \vec{k}) =
g^{\alpha\mu}g^{\beta\nu}-g^{\alpha\nu}g^{\beta\mu},\nonumber\\
&&\eta^{\alpha\beta}(\lambda_1 ,
\lambda'_1,\vec{k})\eta_{\alpha\beta}(\lambda_2 ,
\lambda'_2,\vec{k})=
g_{\lambda_1\lambda_2}g_{\lambda'_1\lambda'_2}-g_{\lambda_1\lambda'_2}g_{\lambda'_1\lambda_2}
\end{eqnarray}
 has been used. The noise
polarization tensor $ K^{\alpha\beta}_N$ defined by (\ref{M8}) can
now be expressed  in terms of the ladder operators of the medium as
\begin{eqnarray}\label{M17}
&&K^{\alpha\beta}_N(x)=\frac{1}{4\pi}\int_0^\infty d\omega\
f^{\alpha\beta\mu\nu}(\omega , x)\sum_{\lambda,\lambda'=0}^3\int
_{-\infty}^{+\infty} d^3k\
\sqrt{\frac{\hbar}{4(2\pi)^3\sqrt{k^2+\omega^2}}}\times\nonumber\\
&&\left[
b_{\lambda\lambda'}(\omega,\vec{k})e^{\imath\vec{k}\cdot\vec{x}-\imath\sqrt{k^2+\omega^2}t}+h.c\right]\eta_{\mu\nu}(\lambda,\lambda',\vec{k})
\end{eqnarray}

In this quantization scheme the Lorentz condition,  $
\partial_\mu A^\mu=0$,   as an operator identity is not compatible to  the commutation relations
(\ref{M10}). Instead only those state vectors $|\psi\rangle$ in the
Hilbert space of the total system are admissible for which the
expectation value of the gauge condition is satisfied, that is $
\langle\psi|
\partial_\mu
A^\mu|\psi\rangle=0$ \cite{13}.\\
\indent It is remarkable that the coupling tensor $
f^{\mu\nu\alpha\beta}(\omega,x) $ satisfying the relation (\ref{M6})
 for an assumed susceptibility tensor $\chi^{\mu\nu\alpha\beta}$
may not be  unique. But the various choices of
$f^{\mu\nu\alpha\beta}(\omega,x) $ satisfying (\ref{M6})for a given
susceptibility tensor are equivalent and do not change the
commutation relations between the electromagnetic field operators.
This becomes more clear if, using (\ref{M16}) and (\ref{M17}), we
compute the commutation relations of the components of the noise
polarization tensor for arbitrary points $x,x'$ in space-time
\begin{equation}\label{M18}
\left[K^{\mu\nu}_N(x)\ ,\
K^{\alpha\beta}_N(x')\right]=\frac{\imath\hbar}{\pi}\left[
\vartheta(x_0-x'_0)\chi^{\mu\nu\alpha\beta}(x,x')-\vartheta(x'_0-x_0)\chi^{\alpha\beta\mu\nu}(x',x)\right].
\end{equation}
These commutation relations  remain unchanged under the various
choices
 of the coupling tensor$f^{\mu\nu\alpha\beta}(\omega,x) $
satisfying (\ref{M6}).\\
\indent For a moving  homogeneous bulk material, that is a medium
for which the tensor $\chi^{\mu\nu\alpha\beta}(x,x')$ is dependent
only on the difference $x-x'$, it is clear  from (\ref{M6})that the
coupling tensor $ f^{\mu\nu\alpha\beta}(\omega,x)$ is independent of
$x$ and the wave equation (\ref{M12}) can be solved using the four
dimensional Fourier transform technique to obtain the vector
potential as
\begin{eqnarray}\label{M19}
&&A_\nu(x)=\int_0^{+\infty}d\omega \int_{-\infty}^{+\infty} d^3k\
\sum_{\lambda,\lambda'=0}^3\sqrt{\frac{\hbar}{2(2\pi)^3\sqrt{k^2+\omega^2}}}\times\nonumber\\
&&\left[ Z_{\nu\alpha\beta}^*(\vec{k},\omega)\
b_{\lambda\lambda'}(\omega,\vec{k})e^{\imath\vec{k}\cdot\vec{x}-\imath\sqrt{k^2+\omega^2}\ t}+h.c\right]\eta^{\alpha\beta}(\lambda,\lambda',\vec{k})\nonumber\\
\\
&&Z_{\nu\alpha\beta}(\vec{k},\omega)= f^{\mu\sigma}_{\
\ \alpha\beta}(\omega)\left[ \imath q_\mu\ L_{\nu\sigma}^{-1}(q)\right]_{q=(\vec{k}\ ,\  \sqrt{k^2+\omega^2})}\nonumber\\
&&
\end{eqnarray}
where
\begin{equation}\label{M20}
L_{\nu\sigma}(q)=\left[ -q_\mu q^\mu \delta_{\nu\sigma}+8\pi\ q^\mu
q^\alpha \underline{\chi}_{\mu\nu\alpha\sigma}(q)\right]
\end{equation}
and $\underline{\chi}^{\mu\nu\alpha\beta}(q)$  is the four
dimensional Fourier transform of the tensor $
\chi^{\mu\nu\alpha\beta}$. Applying a boost transformation on the
susceptibility tensor $\chi^{\mu\nu\alpha\beta}$, one can show that
a medium which is isotropic in its rest frame, becomes an
anisotropic one when it moves with a constant velocity $\vec{v}$. In
this case the electromagnetic field operators are obtained
dependently on the magnitude  and the direction of the velocity of
the moving medium.

\end{document}